\documentclass[doublecol]{epl2} 
\title{Directional interactions and cooperativity between mechanosensitive membrane proteins}
\shorttitle{} %Insert here a short version of the title if it exceeds 70 characters

\author{Christoph A. Haselwandter\inst{1,2} \and Rob Phillips\inst{2}}
\shortauthor{C. A. Haselwandter \etal}

\institute{                    
\inst{1} Department of Physics and Astronomy, University of Southern
California - Los Angeles, California 90089, USA\\
\inst{2} Department of Applied Physics, California Institute of Technology
- Pasadena, California 91125, USA
}

\abstract{While modern structural biology has provided us with a rich and diverse picture of membrane proteins, the biological function of membrane proteins is often influenced by the mechanical properties of the surrounding lipid bilayer. Here we explore the relation between the shape of membrane proteins and the cooperative function of membrane proteins induced by membrane-mediated elastic interactions. For the experimental model system of mechanosensitive ion channels we find that the sign and strength of elastic interactions depend on the protein shape, yielding distinct cooperative gating curves for distinct protein orientations. Our approach predicts how directional elastic interactions affect the molecular structure, organization, and biological function of proteins in crowded~membranes.}

\begin{document}

\maketitle

\section{Introduction}

Cell membranes exhibit a complex organization of lipids and membrane proteins
\cite{engelman05,lang10} and play an integral role in many cellular processes.
The functional properties of membrane proteins are not purely determined by protein structure but, rather, membrane proteins act in concert with the surrounding lipid bilayer \cite{bowie05,brohawn12,milescu09,bosmans11,schmidt12}. In particular, the hydrophobic regions of membrane proteins couple to the hydrophobic regions of lipid bilayers \cite{mouritsen93,mitra04,sonntag11,krepkiy09}. The energetic cost of the resulting membrane deformations can be captured by an elastic model \cite{jensen04,huang86,lundbaek06,andersen07,phillips09,lundbaek10,greisen11}
in which the lipid bilayer is described as an elastic medium and the membrane proteins are regarded as rigid membrane inclusions. Neighboring membrane proteins may induce overlapping deformation fields of the bilayer membrane, yielding long-range interactions between membrane proteins \cite{phillips09,harroun99}.
Such membrane-mediated interactions have been studied using a wide range of analytic and numerical methods
\cite{phillips09,dan93,ursell07,evans03,muller10,yolcu12,reynwar11,goulian93,golestanian96,golestanian96b,west09,brannigan07,aranda96,weikl98,kim98,kim99,kim00,chou01,kim08,lin11},
and experiments have implicated membrane-mediated interactions
in the clustering of a number of different membrane proteins~\cite{harroun99,baddeley09,goforth03,molina06,botelho06,grage11,nomura12}.

Over the past two decades an increasing number of membrane protein structures
have become available \cite{white09,spencer02}, demonstrating a rich diversity in the shapes of membrane proteins. In addition, a variety of experiments have suggested that cell membranes are highly crowded \cite{engelman05,takamori06,dupuy08,linden12}, with the size and spacing of membrane proteins both being of the order of a few nanometers. How are the observed shapes of membrane proteins reflected in the structure of elastic bilayer deformations, and what are the resulting directional interactions between membrane proteins at the small separations most relevant for cell membranes? In this letter we build on the methods in Refs.~\cite{goulian93,dan93,aranda96,weikl98} to develop an analytic approach addressing these questions, and apply this approach to the mechanosensitive channel of large conductance (MscL) \cite{hamill01,markin04,haswell11}. MscL was one of the first membrane ion channels for which high-resolution structural information became available \cite{chang98,sukharev01a,sukharev01b,spencer02,perozo02a,perozo02b} and provides a widely-studied model system for how bilayer mechanical properties regulate protein conformational changes. Our calculation of the interaction potentials and cooperative gating curves for pentameric MscL
\cite{chiang04,anishkin05,chang98,sukharev01a,sukharev01b,spencer02,perozo02a,perozo02b,haswell11,dorwart10,iscla11,gandhi11} shows that, similarly as lipid bilayer material properties 
\cite{nomura12,jensen04,lundbaek06,andersen07,lundbaek10,greisen11,phillips09,perozo02a,perozo02b,chiang04,anishkin05},
directional interactions can alter the structure and function of proteins in crowded membranes.

\section{Calculation of interaction energy}

Following the standard framework of membrane elasticity 
\cite{boal02,safran03,seifert97}, we describe the positions of the inner
and outer lipid bilayer leaflets within the Monge representation through the functions $h_+(x,y)$ and $h_-(x,y)$, which determine \cite{fournier99} the midplane and thickness deformation fields governing bilayer membranes. Since thickness deformations are thought to be dominant for MscL gating \cite{wiggins04,wiggins05,ursell07,ursell08}, we illustrate our method for the thickness deformation field
\begin{equation}
u(x,y)=\frac{1}{2} \left[h_+(x,y)-h_-(x,y)-2a\right]\,, 
\end{equation}
where $2a$ is the hydrophobic thickness of the unperturbed lipid bilayer, with the elastic energy 
\cite{huang86,dan93,boal02,safran03,seifert97,fournier99,nielsen98,dan98,wiggins04,wiggins05,ursell07,ursell08}
\begin{equation} \label{eq1}
{\textstyle G=\frac{1}{2}}\int dx dy {\textstyle\left\{K_b (\nabla^2 u)^2+K_t \left(\frac{u}{a}\right)^2+\tau
\left[2 \frac{u}{a}+(\nabla u)^2 \right] \right\}}\,,
\end{equation}
where $K_b$ is the bending rigidity of the lipid bilayer, $K_t$ is the stiffness associated with thickness deformations, and $\tau$ is the membrane tension.
Midplane deformations decouple to leading order from thickness deformations
\cite{fournier99}, and are described \cite{boal02,safran03,seifert97,fournier99,wiggins04,wiggins05,ursell07,ursell08} by an energy functional similar to
eq.~(\ref{eq1}). For generality we allow for the two tension terms $u/a$ and $\left(\nabla u\right)^2$ in eq.~(\ref{eq1}), which capture the effects of membrane tension on lipid surface area \cite{safran03,ursell07,ursell08} and on membrane undulations \cite{boal02,safran03,seifert97,fournier99,wiggins04,wiggins05}, respectively.

The Euler-Lagrange equation associated with eq.~(\ref{eq1}) is given by
\begin{equation} \label{eq2}
\left(\nabla^2-\nu_+\right) \left(\nabla^2-\nu_-\right) \bar u =0\,,
\end{equation}
where 
\begin{equation}
\nu_\pm = \frac{1}{2 K_b}\left[\tau \pm \left(\tau^2-\frac{4 K_b K_t}{a^2} \right)^{1/2}\right]
\end{equation}
and $\bar u = u + \tau a/K_t$. The solution of eq.~(\ref{eq2}) for a single cylindrical membrane inclusion of radius $R_i$ is \cite{zauderer83,huang86}
\begin{equation} \label{eq3}
\bar u(r_i,\theta_i) = f^+_i(r_i,\theta_i) + f^-_i(r_i,\theta_i)\,,
\end{equation}
where $r_i$ and $\theta_i$ are polar coordinates with the center of inclusion
$i$ as the origin, the Fourier-Bessel series 
\begin{equation}
f^{\pm}_i(r_i,\theta_i)= A_{i,0}^\pm K_0(\sqrt{\nu_\pm} r_i) +\sum_{n=1}^\infty \left(\mathcal{A}_{i,n}  +\mathcal{B}_{i,n} \right)\,,
\end{equation}
in which $\mathcal{A}_{i,n}= A_{i,n}^\pm K_n(\sqrt{\nu_\pm} r_i) \cos n \theta_i$ and $\mathcal{B}_{i,n}= B_{i,n}^\pm K_n(\sqrt{\nu_\pm} r_i)\sin n \theta_i$, $K_n$ are modified Bessel functions of the second kind,  and we have assumed that membrane deformations decay away from a single membrane inclusion \cite{nielsen98}.
The coefficients $A_{i,n}^\pm$ and $B_{i,n}^\pm$ are determined by the boundary conditions at the bilayer-inclusion interface for which, to allow for general protein shapes, we~use
\begin{eqnarray} \label{genBC1}
u(r_i,\theta_i)\big|_{r_i=R_i}&=&U_i(\theta_i)\,, \\ \label{genBC2}
\mathbf{\hat n} \cdot \nabla u(r_i,\theta_i)\big|_{r_i=R_i}&=&U_i^\prime(\theta_i)\,,
\end{eqnarray}
where $\mathbf{\hat n}$ is the unit normal vector along the bilayer-inclusion
interface. As an alternative to eq.~(\ref{genBC2}) one may assume \cite{aranda96,dan93,nielsen98,brannigan06,brannigan07} a free contact slope along the bilayer-inclusion interface.

Following Refs.~\cite{goulian93,weikl98}, we construct the solution of eq.~(\ref{eq2})
for two membrane inclusions using the ansatz
\begin{equation} \label{genSol}
u=u_1(r_1,\theta_1)+u_2(r_2,\theta_2)\,,
\end{equation}
where the $u_i(r_i,\theta_i)$ are the single-inclusion deformation fields implied by eq.~(\ref{eq3}), with the bipolar coordinate
transformations 
\begin{equation}
r_2=\left(d^2+r_1^2+2 d r_1 \cos \theta_1 \right)^{1/2}\,,
\end{equation}
$\cos \theta_2 = \left(d+r_1 \cos \theta_1\right)/r_2$, and $\sin \theta_2 = \left(r_1 \sin \theta_1\right)/r_2$, in which $d$ is the center-to-center distance between the two inclusions. Note that, if the single-inclusion solution in eq.~(\ref{eq3}) is considered up to some order $n=N$, eq.~(\ref{genSol}) contains $4(2N+1)$ independent constants which must be fixed through the boundary conditions in eqs.~(\ref{genBC1}) and~(\ref{genBC2}). Employing eq.~(\ref{eq2}), the elastic energy in eq.~(\ref{eq1}) can then be written as a sum over the boundary terms
\begin{equation} \label{eq4}
G_i=- \frac{R_i}{2} \int_0^{2 \pi} d \theta_i \left(K_b\frac{\partial \bar u}{\partial r_i} \nabla^2 \bar u-K_b \bar u  \frac{\partial}{\partial
r_i}\nabla^2 \bar u +\tau \bar u \frac{\partial \bar u}{\partial r_i} \right)
\end{equation}
at $r_i=R_i$, where we have neglected terms independent of $\bar u$. This expression is evaluated at each order in the Fourier-Bessel series in eq.~(\ref{genSol}) using the orthogonality properties of trigonometric functions.
Thus, the elastic energy in eq.~(\ref{eq1}) reduces to an algebraic
expression for the coefficients $A_{i,n}^\pm$ and $B_{i,n}^\pm$ which, in turn, are prescribed by the boundary conditions in eqs.~(\ref{genBC1}) and~(\ref{genBC2}).

We determine the coefficients $A_{i,n}^\pm$ and $B_{i,n}^\pm$ in eq.~(\ref{genSol}) up to arbitrary $N$ by expanding $u_2$ in the membrane region surrounding inclusion 1 in terms of $r_1/d<1$ and imposing on $u$ the boundary conditions at $r_1=R_1$ \cite{goulian93,weikl98}, with a similar procedure for inclusion 2. Thus, finding the general solution in eq.~(\ref{genSol}) which respects the boundary conditions in eqs.~(\ref{genBC1}) and~(\ref{genBC2}) reduces to solving a system of $4(2N+1)$ linear equations, which we achieve using standard methods \cite{mathematica}. Substitution of the resulting expressions of the coefficients into eq.~(\ref{eq4}) yields, for a given $N$, the interaction potential between two membrane inclusions. The validity of this finite-order series solution is based on the assumption that very rapid angular variations at large $n$ in the membrane deformation field can be neglected, which we confirm for a given problem by systematically including higher-order terms.

\section{Cylinder model of MscL}

Before employing eq.~(\ref{eq4}) with eqs.~(\ref{genBC1}) and~(\ref{genBC2}) to study directional interactions between MscL proteins we test our approach
in the special case of membrane inclusions with circular cross section, for which elastic interactions have been investigated in some detail
\cite{ursell07,phillips09,dan93,muller10,yolcu12,reynwar11,goulian93,west09,brannigan07,aranda96,weikl98,kim98,kim99,chou01,evans03}.
In particular, we consider cylindrical membrane inclusions of constant hydrophobic thickness, which have been used to model MscL \cite{grage11,wiggins04,wiggins05,ursell07,ursell08} as well as other membrane proteins \cite{harroun99,jensen04,huang86,lundbaek06,andersen07,phillips09,lundbaek10,greisen11}.
Following the basic phenomenology of MscL gating \cite{hamill01,markin04,haswell11}, we take the inclusion to exist in one of two states---open or closed---with the competition between these two states governed by membrane tension. Furthermore, in the cylinder model of MscL \cite{grage11,wiggins04,wiggins05,ursell07,ursell08}, open and closed states of MscL are distinguished only by the inclusion radius and hydrophobic thickness.

\begin{figure}[t!]
\onefigure{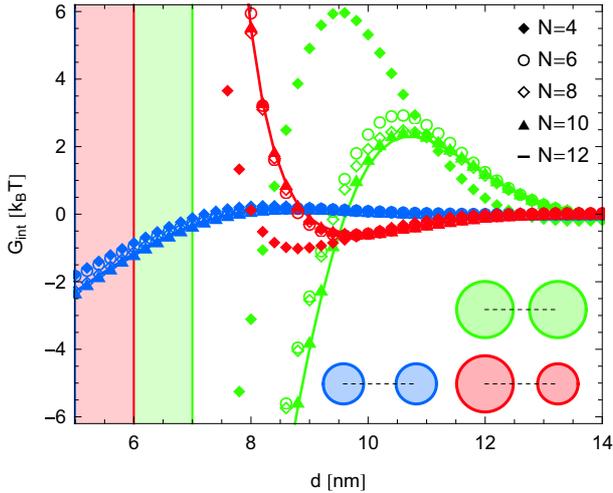}
\caption{\label{fig1} (Color on-line) Elastic interaction potentials obtained from eq.~(\ref{eq4}) between cylindrical membrane inclusions of constant hydrophobic thickness up to order $n=N$ in the Fourier-Bessel series in eq.~(\ref{genSol}) between open, closed, and open and closed channel states (curves from top to bottom at $d=10$~nm) for $\tau=0$. We use the same parameter values as
in Ref.~\cite{ursell07}. Vertical lines and shaded regions indicate the minimum values of $d$ mandated by steric constraints.}
\end{figure}

Figure~\ref{fig1} shows our solutions of the elastic interaction potentials for cylindrical membrane inclusions obtained from eq.~(\ref{eq4}) at various $N$. Because we expand the solutions in $r_i/d$ around each inclusion $i$, convergence is slowest at small $d$. The interaction potentials in fig.~\ref{fig1} exhibit the same qualitative behavior as observed in computer simulations of this system \cite{ursell07,grage11}. Moreover, for intermediate and large $d$, the results in fig.~\ref{fig1} are in good quantitative agreement with the corresponding numerical solutions of the elastic equations \cite{ursell07}. For small $d$, however, the values of the interaction potentials for closed (and open) channels obtained from simulations \cite{ursell07} are of a somewhat smaller magnitude than those in fig.~\ref{fig1} at $N=8$ and beyond, while smaller values of $N$ also produce smaller magnitudes of the interaction energy in our analytic calculation. Thus, a potential explanation for the discrepancy between analytic and numerical results at small~$d$ is that the grid size
used for the simulations in Ref.~\cite{ursell07} does not capture rapid angular variations (which, in our analytic solution, correspond to large values of $N$) with sufficient accuracy at small inclusion separations.

\section{Pentameric MscL}

\begin{figure}[t!]
\onefigure{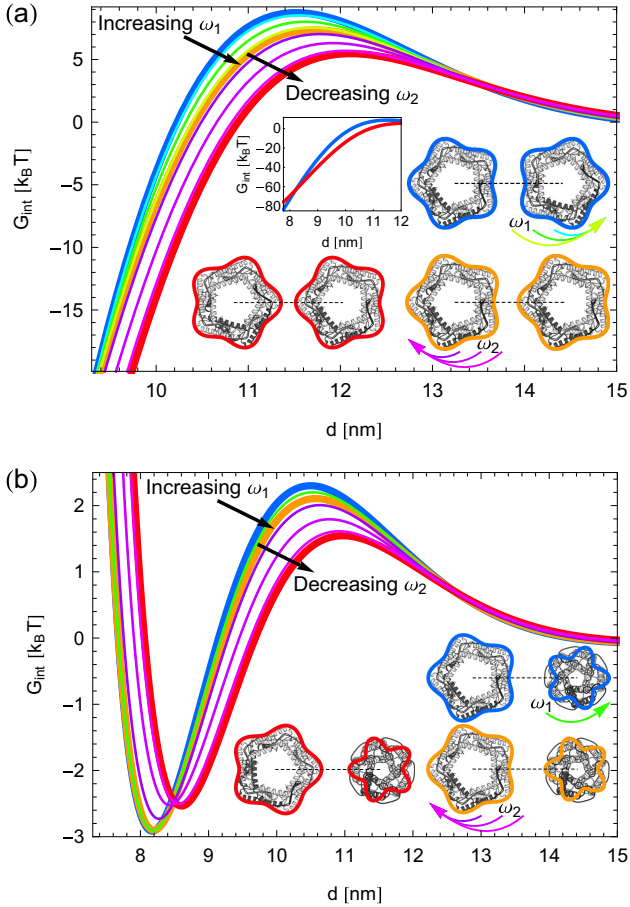}
\caption{\label{fig2}(Color on-line) Directional interaction potentials obtained
from eq.~(\ref{eq4}) at $N=12$ for (a) open and (b) open and closed MscL. Bilayer-MscL interactions were parameterized as in Ref.~\cite{ursell08} but for a PC20 lipid bilayer with $\tau=0$. Thick curves denote the three MscL orientations in the insets, while thin curves correspond to intermediate MscL orientations rotated by $\pi/20$ or $\pi/10$ as indicated by arrows. The molecular models of MscL in the insets are reprinted, with permission, from \textit{Annual Review of Biophysics and Biomolecular Structure} (Volume 31 $\copyright$ 2002 by Annual Reviews; www.annualreviews.org; ref.~\cite{spencer02}),
and the superimposed boundary curves were obtained from eq.~(\ref{boundC}) with $R_i\approx3.49$~nm and $R_i\approx2.27$~nm, and $\epsilon_i\approx0.11$ and $\epsilon_i\approx0.22$, for open and closed~MscL.}
\end{figure}

While the cylinder model of membrane proteins \cite{jensen04,huang86,lundbaek06,andersen07,phillips09,lundbaek10,greisen11}
provides a beautiful zeroth-order description of the membrane deformation footprint of a membrane protein, structural biology has shown that membrane
protein structures are much richer. In particular, the structures of MscL
in different organisms have been found to exhibit distinct symmetries 
\cite{haswell11,dorwart10,iscla11,gandhi11}. One of the primary goals of the work described here is to determine in what way the zeroth-order model of membrane-mediated protein interactions needs to be amended when accounting for real membrane proteins.
A simple coarse-grained model of the cross-sectional shape of MscL, as well
as of other membrane proteins, is given by
\begin{equation} \label{boundC}
C_{i}(\theta_i)=R_i \left[1+\epsilon_i \cos s \left(\theta_i-\omega_i\right) \right]\,,
\end{equation}
where $\epsilon_i$ parameterizes the magnitude of the deviation of the protein cross section from the circle, $s$ denotes the order of the protein
symmetry (oligomeric state), and $\omega_i$ captures the orientation of the protein. Equation~(\ref{boundC}) is illustrated in the insets of fig.~\ref{fig2}
for open and closed MscL.

The physiologically relevant oligomeric states of MscL remain a matter of debate \cite{haswell11,dorwart10,iscla11,gandhi11}, with tetrameric \cite{liu09}, pentameric \cite{chang98}, and hexameric \cite{saint98} MscL having been reported. Our approach is able to handle all of these cases, but here
we focus on pentameric MscL, which correspond to $s=5$ in eq.~(\ref{boundC}) with $R_i$ and $\epsilon_i$ estimated from structural models \cite{chang98,sukharev01a,sukharev01b,spencer02,perozo02a,perozo02b,chiang04,anishkin05}.
Since these structural models of MscL suggest $\epsilon_i\ll1$, we take the weak perturbation limit of eq.~(\ref{boundC}) and only consider the
leading-order terms in $\epsilon_i$ breaking rotational symmetry. Expansion of $u$ at $r_i=C_{i}(\theta_i)$ in $\epsilon_i$ \cite{kim00,kim08} then yields boundary conditions of the form in eqs.~(\ref{genBC1}) and~(\ref{genBC2})
which, following the procedure described above, allows us to analytically solve for the directional interaction potentials between
MscL proteins. For simplicity we first consider the case of a constant $U_i$ along $C_{i}(\theta_i)$, with a value of $a$ so that $U_i$ is of the same sign in the open and closed states of MscL, and then turn to the complementary case of directional interactions induced by a varying~$U_i$ \cite{sonntag11,krepkiy09}. We choose the parameter values for bilayer-MscL interactions as discussed in Refs.~\cite{wiggins04,wiggins05,ursell07,ursell08} and, in particular, set $U_i^\prime(\theta_i)=0$ along the bilayer-MscL~interface.

Figure~\ref{fig2}(a) shows the elastic interaction potentials between pentameric MscL proteins in the open state \cite{sukharev01a,sukharev01b,perozo02b,perozo02a,spencer02}, with similar curves for the corresponding closed state structure \cite{chang98}. Irrespective of the relative protein orientation considered, we find an energy barrier to the dimerization of two open (or closed) MscL proteins. For most protein separations, the tip-on orientation is energetically most favorable with the face-on orientation being least favorable. For very small $d$, however, the face-on orientation becomes favorable over the tip-on orientation [see inset in fig.~\ref{fig2}(a)]. Figure~\ref{fig2}(b) demonstrates that elastic interactions between open and closed MscL proteins can also yield an energy barrier to dimerization. But, in contrast to fig.~\ref{fig2}(a), MscL proteins now repel each other at very small $d$, and we obtain a pronounced minimum in the interaction energy at some optimal value of $d$ which depends on the relative protein orientation. Similarly as in the case of two open MscL proteins, the tip-on orientation is energetically most favorable, and the face-on orientation least favorable, for most protein separations in fig.~\ref{fig2}(b), with the tip-on orientation becoming least favorable for very small $d$. For all scenarios in fig.~\ref{fig2} we find that $G$ changes smoothly upon rotation from the face-on to the tip-on orientation.

The analytic solution in eq.~(\ref{genSol}) in terms of single-inclusion
deformation fields suggests a simple qualitative explanation for the non-monotonic behavior of the MscL interaction potentials in fig.~\ref{fig2}. In general, the single-inclusion thickness deformation field in eq.~(\ref{eq3})
overshoots considerably \cite{huang86,dan93} at its first extremum when relaxing away from the inclusion
boundary. Hence, if two channels induce thickness deformations of the same sign, there is a regime of intermediate $d$ for which membrane deformations are amplified, yielding the energy barriers in fig.~\ref{fig2}. At small $d$, the membrane deformation fields already overlap before reaching the first extremum, thus reducing the overall deformation footprint of the two channels and making dimerization favorable as implied by fig.~\ref{fig2}. For membrane channels of distinct hydrophobic thickness, however, moving the channels even closer together yields an
additional regime in which the membrane has to deform strongly between the two channels, leading to repulsion between open and closed MscL at very small $d$ as in fig.~\ref{fig2}(b). In each of these regimes, the competition between tip-on and face-on orientations is governed by a complex interplay between the local strengths and associated membrane areas of overlapping deformation~fields.

\section{Varying hydrophobic thickness}

\begin{figure}[t!]
\onefigure{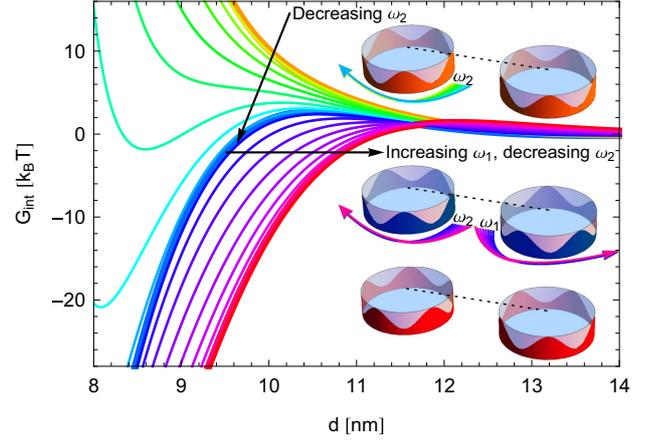}
\caption{\label{fig3}(Color on-line) Directional interaction potentials obtained
from eq.~(\ref{eq4}) at $N=12$ for cylindrical membrane inclusions of radius $R_i=3.5$~nm \cite{ursell07} with the hydrophobic mismatch in eq.~(\ref{VarU}) using $U_i^0=-0.5$~nm \cite{ursell07} and $\delta_i=0.7$~nm. The lipid bilayer was parameterized as in Ref.~\cite{ursell07} with $\tau=0$. Thick curves denote the three inclusion orientations in the insets, while thin curves correspond to intermediate orientations rotated by $\pi/45$ as indicated by arrows.}
\end{figure}

The qualitative picture of elastic interaction potentials developed above for the boundary curves in eq.~(\ref{boundC}) also applies to the complementary
case of directional interactions induced by a varying hydrophobic thickness
of membrane proteins \cite{sonntag11,krepkiy09}. To explore this scenario we consider cylindrical membrane inclusions with hydrophobic mismatch
\begin{equation} \label{VarU}
U_{i}(\theta_i)=U_i^0+ \delta_i \cos s \left(\theta_i-\omega_i\right)\,,
\end{equation}
where $U_i^0$ is the average hydrophobic mismatch and $\delta_i$ is the
magnitude of mismatch modulations. The hydrophobic surfaces generated by eq.~(\ref{VarU}) are illustrated in the insets of fig.~\ref{fig3}.

Figure~\ref{fig3} shows the interaction potentials associated with eq.~(\ref{VarU}). A striking feature of fig.~\ref{fig3} is that, depending on the relative inclusion orientation, membrane-mediated interactions can be attractive or repulsive at small $d$. In particular,
if the periodic modulations in eq.~(\ref{VarU}) are in phase at the closest approach of the two inclusions there is, again due to the oscillatory nature of the single-inclusion deformation fields, an energy barrier to dimerization as in fig.~\ref{fig2}(a), with attraction at small $d$. Conversely, in the case of orientations in fig.~\ref{fig3} for which the periodic modulations in eq.~(\ref{VarU}) are out of phase at the closest approach of the two inclusions, opposing inclusion boundaries induce distinct thickness deformations which, similarly as in fig.~\ref{fig2}(b), can lead to repulsion at small $d$. The interaction potentials in fig.~\ref{fig3} change smoothly upon rotation from out-of-phase to in-phase orientations, and exhibit a minimum for intermediate orientations.

\section{Cooperative gating}

Interaction potentials such as those in figs.~\ref{fig2} and~\ref{fig3} can
induce directionality in the cooperative function of membrane proteins. The gating characteristics of MscL with varying membrane
tension are captured \cite{hamill01,markin04}
by the channel opening probability
\begin{equation} \label{gatingP}
P_o=\frac{1}{1+e^{\beta \left(\Delta G-\tau \Delta A\right)}}\,,
\end{equation}
where $\beta=1/k_B T$, in which $k_B$ is Boltzmann's constant and $T$ is
the temperature, and $\Delta G$ and $\Delta A$ are the free energy and area difference between open and closed channel states. We only consider here
contributions to $\Delta G$ due to thickness deformations of the bilayer
membrane in eq.~(\ref{eq4}), which were shown previously \cite{wiggins04,wiggins05,ursell07,ursell08,grage11} to yield the basic phenomenology of MscL gating.

\begin{figure}[t!]
\onefigure{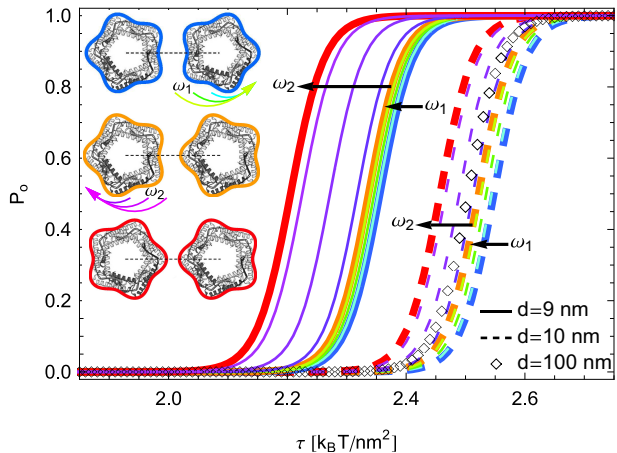}
\caption{\label{fig4} (Color on-line) Gating curves in eq.~(\ref{gatingP}) for a pair of MscL proteins to transition from the open-closed to the open-open configuration [see figs.~\ref{fig2}(b) and~\ref{fig2}(a)] for the lipid bilayer parameter values in Ref.~\cite{ursell07} at $d=9$~nm, $d=10$~nm, and the
value $d=100$~nm corresponding to the far-field limit. All gating curves are obtained from eq.~(\ref{eq4}) at $N=12$, and the protein orientations at $d=9$~nm and $d=10$~nm
are labelled as in fig.~\ref{fig2}(a). The molecular models of MscL in the insets are reprinted, with permission, from \textit{Annual Review of Biophysics and Biomolecular Structure} (Volume 31 $\copyright$ 2002 by Annual Reviews; www.annualreviews.org; ref.~\cite{spencer02}), and we use the same MscL shapes as in fig.~\ref{fig2}.}
\end{figure}

Figure~\ref{fig4} shows $P_o$ for pentameric MscL 
\cite{chiang04,anishkin05,chang98,sukharev01a,sukharev01b,spencer02,perozo02a,perozo02b,haswell11,dorwart10,iscla11,gandhi11} with a neighboring pentameric MscL protein in the open state. We find that, compared to the case of non-interacting MscL, membrane-mediated interactions can, depending on the protein separation and orientation, shift the gating tension to higher as well as lower values. The magnitude of the predicted
effect of directional interactions in fig.~\ref{fig4} is comparable to previously measured \cite{perozo02b,nomura12} shifts in gating tension due to modification of the membrane composition. As expected from the interaction potentials in fig.~\ref{fig2}, the cooperative gating tension associated with the tip-on orientation is lower than the cooperative gating tension associated with the face-on orientation for the protein separations in fig.~\ref{fig4}, with a smooth interpolation in $P_o$ between these two limiting orientations.

\section{Conclusion}

Building on Refs.~\cite{goulian93,dan93,aranda96,weikl98} we have developed an analytic approach for estimating directional elastic interactions between membrane proteins for arbitrary protein separations. On the basis of a simple model of the shape of MscL suggested by structural studies
\cite{anishkin05,chiang04,haswell11,dorwart10,iscla11,gandhi11,chang98,sukharev01a,sukharev01b,spencer02,perozo02a,perozo02b},
we predict that directional interactions between MscL proteins yield a characteristic sequence of preferred orientations, and cooperative gating curves, upon dimerization of MscL. A combination of quantitative experiments on the spatial arrangement \cite{grage11,nomura12} and gating tension \cite{markin04,perozo02a,perozo02b,chiang04,anishkin05} of MscL would be able to put these predictions to a direct experimental test.
Our method provides a bridge connecting the shape of membrane proteins to the cooperative function of membrane proteins induced by elastic interactions, and can be applied to any membrane protein for which basic structural information is available.  The approach developed here represents a step towards a physical theory of how directional interactions affect the molecular structure, organization, and biological function of proteins in the crowded membrane environment provided by living cells.

\acknowledgments

This work was supported at USC by the National Science Foundation through
NSF award number DMR-1206332 and at Caltech by a Collaborative Innovation Award of the Howard Hughes Medical Institute, and the National Institutes of Health through NIH award number R01 GM084211 and the Director's Pioneer Award. We thank C. L. Henley, W. S. Klug, M. Lind\'en, D. C. Rees, and N.~S. Wingreen for helpful comments.

\end{document}